\documentclass[showpacs,prl,floatfix,twocolumn,amsmath,a4]{revtex4}
\usepackage{graphicx}
\usepackage{amsmath}
\usepackage{amssymb}
\usepackage{bm}
\usepackage{dcolumn}

\begin{document}
\title{Doping graphene with metal contacts}
\author{G. Giovannetti$^{1,2}$, P. A. Khomyakov$^2$, G. Brocks$^2$, V. M. Karpan$^{2}$,
J. van den Brink$^{1,3}$, and P. J. Kelly$^{2}$}
\affiliation{$^{1}$Instituut-Lorentz for Theoretical Physics,
Universiteit Leiden, P.O. Box 9506, 2300 RA Leiden, The Netherlands\\
$^{2}$Faculty of Science and Technology and MESA$^+$ Institute for
Nanotechnology, University of Twente, P.O. Box 217, 7500 AE Enschede,
The Netherlands.\\
$^{3}$Institute for Molecules and Materials, Radboud Universiteit, Nijmegen, The Netherlands. }

\begin{abstract}
Making devices with graphene necessarily involves making contacts with
metals. We use density functional theory to study how graphene is doped
by adsorption on metal substrates and find that weak bonding on Al, Ag,
Cu, Au and Pt, while preserving its unique electronic structure, can
still shift the Fermi level with respect to the conical point by $\sim
0.5$ eV. At equilibrium separations, the crossover from $p$-type to
$n$-type doping occurs for a metal work function of $\sim 5.4$ eV, a
value much larger than the graphene work function of 4.5 eV. The
numerical results for the Fermi level shift in graphene are described
very well by a simple analytical model which characterizes the metal
solely in terms of its work function, greatly extending their
applicability.
\end{abstract}
\date{\today}
\pacs{73.63.-b, 73.20.Hb, 73.40.Ns, 81.05.Uw}
\maketitle

Recent progress in depositing a single graphene sheet on an insulating
substrate by micromechanical cleavage enables electron transport
experiments on this two-dimensional system
\cite{Novoselov:sc04,Novoselov:nat05}. Such experiments demonstrate an
exceptionally high electron mobility in graphene, quantization of the
conductivity, and a zero-energy anomaly in the quantum Hall effect, in
agreement with theoretical predictions
\cite{Shon:jpsj98,Ando:jpsj02,Gusynin:prl05,Katsnelson:natp06,Brink:natn07}.
The spectacular effects arise from graphene's unique electronic
structure. Although it has a zero band gap and a vanishing density of
states at the Fermi energy, graphene exhibits metallic behavior due to
topological singularities at the $K$-points in the Brillouin zone
\cite{Shon:jpsj98,Ando:jpsj02} where the conduction and valence bands
touch in
conical (Dirac) points
and the dispersion is
essentially linear within $\pm 1$ eV of the Fermi energy.

In a free-standing graphene layer the Fermi energy coincides with the
conical points but adsorption on metallic (or insulating) substrates
can alter its electronic properties significantly
\cite{Oshima:jpcm97,Dedkov:prb01,Bertoni:prb05,NDiaye:prl06,Karpan:prl07,Giovannetti:prb07,Marchini:prb07,Uchoa:prb08}.
Since electronic transport measurements through a graphene sheet require contacts to metal electrodes
\cite{Novoselov:nat05,Karpan:prl07,Schomerus:prb07,Blanter:prb07} it is
essential to have a full understanding of
the physics of metal-graphene interfaces. In this
paper we use first-principles calculations at the level of density
functional theory (DFT) to study the adsorption of graphene on a series
of metal substrates. The (111) surfaces of Al, Co, Ni, Cu, Pd, Ag, Pt
and Au, covering a wide range of work functions and chemical bonding,
form a suitable system for a systematic study.

\begin{figure}[bp]
\includegraphics[width=0.9\columnwidth,angle=0]{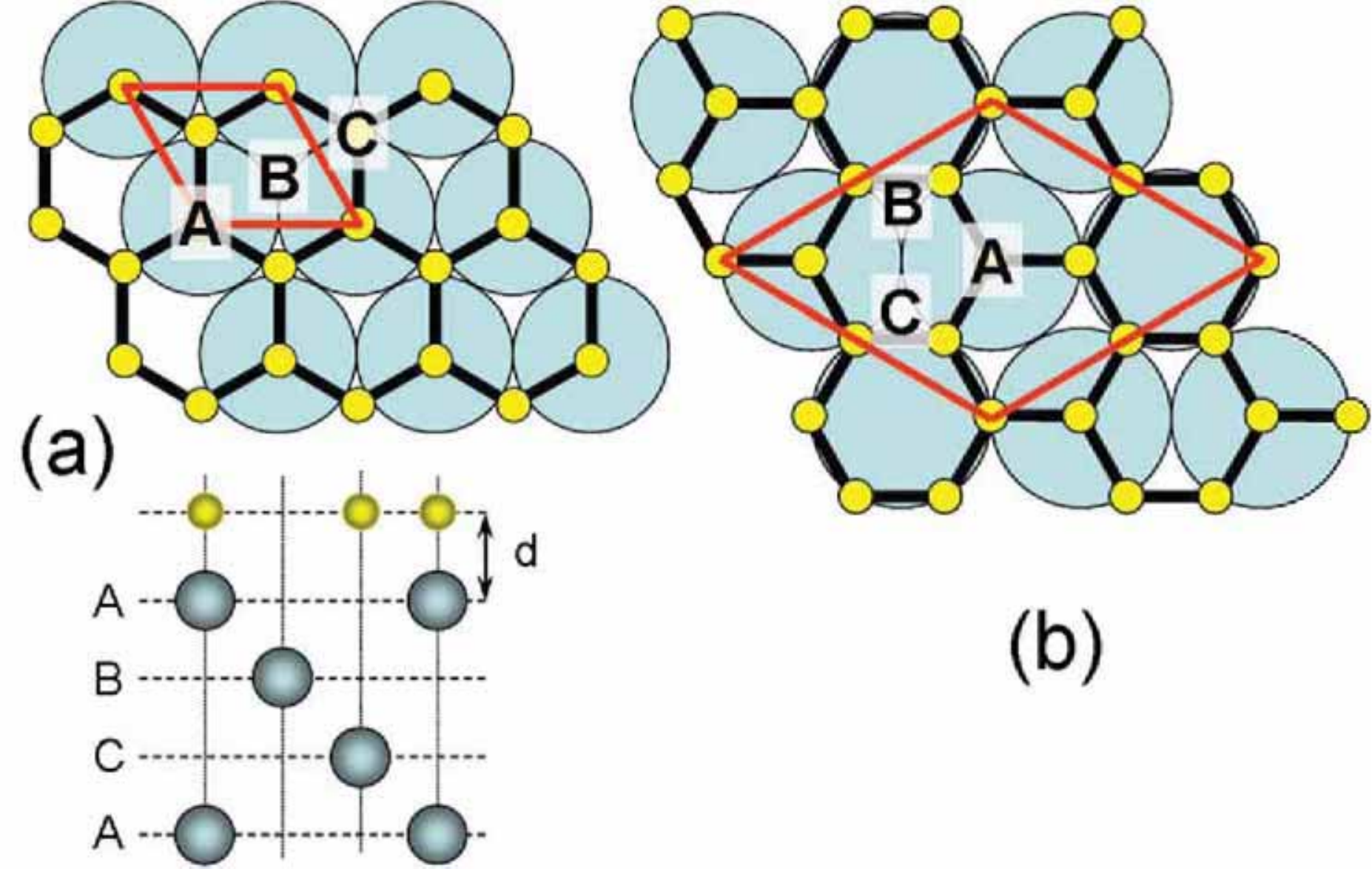}
\caption{(Color online) The most stable configurations of graphene
(a) on Cu,  Ni and Co (111) with one carbon atom on top of a metal atom
(A site), and the second carbon on a hollow site (C site) and
(b) on Al, Au, Pd and Pt(111) in a unit cell with
8 carbon atoms and 3 metal atoms per layer.}
\label{ref:fig1}
\end{figure}

Our results show that these substrates can be divided into two classes.
The characteristic electronic structure of graphene is significantly
altered by chemisorption on Co, Ni and Pd but is preserved by weak
adsorption on Al, Cu, Ag, Au and Pt. Even when the bonding is weak,
however, the metal substrates cause the Fermi level to move away from
the conical points in graphene, resulting in doping with either
electrons or holes. The sign and amount of doping can be deduced from
the difference of the metal and graphene work functions only when they
are so far apart that there is no wave function overlap. At the
equilibrium separation, the doping level is strongly affected by an
interface potential step arising from the direct metal-graphene
interaction.

Based upon the DFT results, we develop a phenomenological model to
describe the doping of graphene, taking into account the metal-graphene
interaction. The model uses only the work functions of graphene and of
the clean metal surfaces as input. For a given metal substrate, it
allows us to predict the Fermi level shift in graphene with respect to
the conical points {\em i.e.,} both the type and concentration of the
charge carriers. The model also predicts how metal work functions are
modified by adsorption of graphene.

\begin{figure}[!tpb]
\begin{center}
\includegraphics[width=1.0\columnwidth]{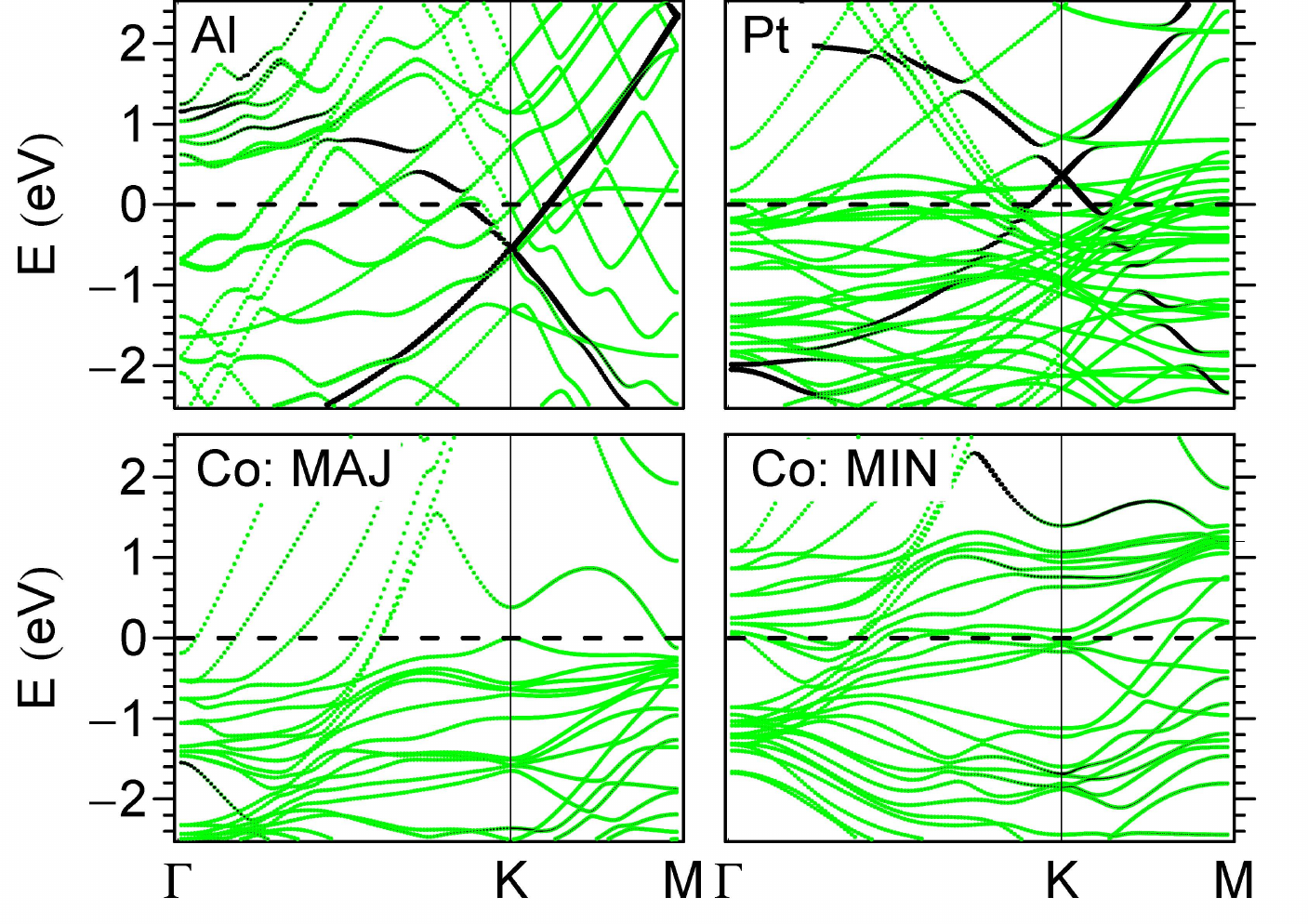}
\end{center}
\caption{(Color online) Band structures of graphene absorbed upon Al,
Pt and Co (111) substrates. The Fermi level is at zero energy. The
amount of carbon $p_{z}$ character is indicated by the blackness of the
bands. The conical point corresponds to the crossing of predominantly
$p_{z}$ bands at $K$.
Top panels: graphene on Al and Pt; bottom panels:
the majority and minority spin bands of graphene on Co.
Note that on
doubling the lattice vectors (for Al and Pt), the $K$ point is folded
down onto the $K$ point of the smaller Brillouin zone. }
\label{ref:fig2}
\end{figure}

\begin{table}[b]
\caption{Calculated equilibrium separation $d_{\rm eq}$ of a graphene
sheet from various metal (111) surfaces. The binding energy $\Delta E$
is the energy (per carbon atom) required to remove the graphene sheet
from the metal surface. $W_{\rm M}$ and $W$ are, respectively, the work
functions calculated for the clean metal surfaces, and for free-standing
and adsorbed graphene.}
\begin{ruledtabular}
\begin{tabular}{lccccccccc}
                      & Gr   &   Ni   & Co   & Pd   & Al   &  Ag  & Cu   & Au   & Pt     \\
\hline
 ${d_{\rm eq}}$ (\AA) &      &  2.05  & 2.05 & 2.30 & 3.41 & 3.33 & 3.26 & 3.31 & 3.30  \\
 ${\Delta}E$ (meV)    &      &  125   &  160 & 84   &  27  &  43  & 33   & 30   & 38    \\
 $W_{\rm M}$ (eV)     &      &  5.47  & 5.44 & 5.67 & 4.22 & 4.92 & 5.22 & 5.54 & 6.13  \\
 $W$ (eV)             & 4.48 &  3.66  & 3.78 & 4.03 & 4.04 & 4.24 & 4.40 & 4.74 & 4.87  \\
$W_{\rm exp}$ (eV)    & $4.6^{\rm a}$
                             & 3.9\footnote{Ref.~\cite{Oshima:jpcm97}}
                                      &      &   $4.3^{\rm a}$   &      &      &      &
                                                           & $4.8^{\rm a}$ \label{ref:tab1}
\end{tabular}
\end{ruledtabular}
\end{table}

Some details of how DFT ground state energies and optimized geometries
are calculated for graphene on metal (111) surfaces are given in
Ref.~\cite{compdetails}. We fix the in-plane lattice constant of
graphene to its optimized value $a = 2.445$ \AA\ and adapt the lattice
constants of the metals accordingly. The graphene honeycomb lattice
then matches the triangular lattice of the metal (111) surfaces in the
unit cells shown in Fig.~\ref{ref:fig1}. The approximation made by this
procedure is reasonable, since the mismatch with the optimized metal
lattice parameters is only 0.8-3.8\%. We have verified explicitly that
the structures shown in Fig.~\ref{ref:fig1} represent the most stable
configurations of graphene on the metal substrates studied. The
equilibrium
separations, binding energies and work functions are listed in
Table~\ref{ref:tab1}.

The results immediately show that the metals can be divided into two
classes. Graphene is chemisorbed on Co, Ni and Pd(111), leading to
binding energies $\Delta E\sim 0.1$ eV/carbon atom and equilibrium
separations $d_{\rm eq} \lesssim 2.3$ \AA. In contrast, adsorption on
Al, Cu, Ag, Au and Pt(111) leads to a weaker bonding, $\Delta E
\lesssim 0.04$ eV/carbon atom, and larger equilibrium separations,
$d_{\rm eq} \sim 3.3$ \AA.
These results are in agreement with previous calculations and experimental data
\cite{Oshima:jpcm97,Bertoni:prb05,Karpan:prl07,Gamo:ss97,Qi:ss05}.

\begin{figure}[tbp]
\begin{center}
\includegraphics[width=1.0\columnwidth]{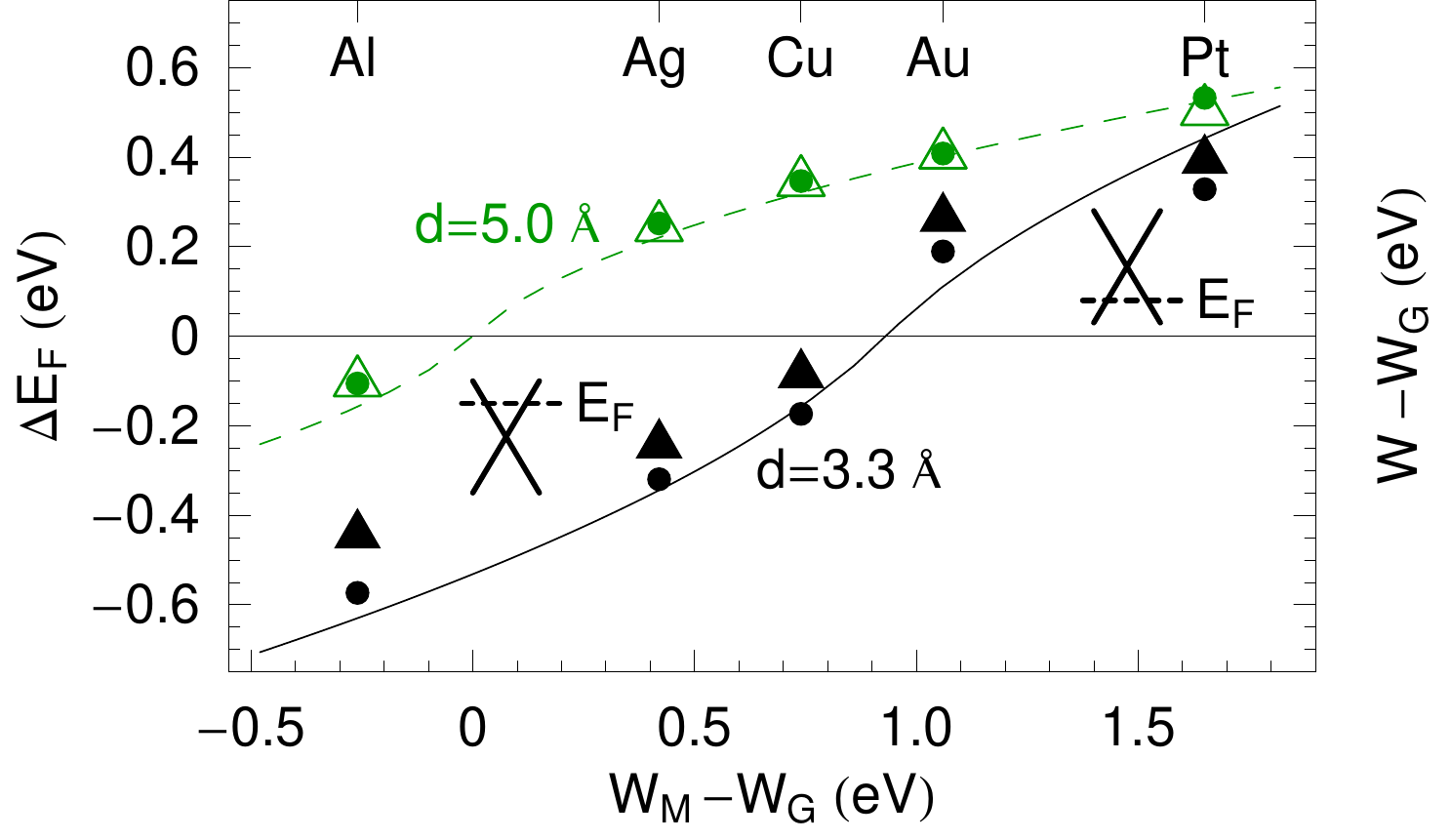}
\caption{(Color online) Calculated Fermi energy shift with respect to
the conical point, $\Delta E_{\rm F}$ (dots), and change in the work
function $W - W_{\rm G}$ (triangles) as a function of $W_{\rm M} -
W_{\rm G}$, the difference between the clean metal and graphene work
functions. The lower (black) and the upper (green/grey) results are for
the equilibrium ($\sim 3.3$ \AA) and a larger (5.0 \AA) separation of
graphene and the metal surfaces, respectively. The solid and dashed
lines follow from the model of Eq.~(\ref{eqn1}) with $\Delta_{\rm c}=0$
for $d=5.0$ \AA. The insets illustrate the position of the Fermi level
with respect to the conical point.} \label{ref:fig3}
\end{center}
\end{figure}

To identify the changes in the graphene electronic structure induced
by adsorption, we calculate the band structures as illustrated in
Fig.~\ref{ref:fig2} for some typical examples. When graphene is
chemisorbed (on Co, Ni, and Pd) the graphene bands are strongly
perturbed and acquire a mixed graphene-metal character. In particular,
the characteristic conical points at $K$ are destroyed, see the bottom
panels of Fig.~\ref{ref:fig2}. When the interaction is weaker (Al, Cu,
Ag, Au, Pt), the graphene bands, including their conical points at $K$,
can still be clearly identified; see the upper panels of
Fig.~\ref{ref:fig2}. However, whereas in free-standing graphene the
Fermi level coincides with the conical point, adsorption generally
shifts the Fermi level. A shift upwards (downwards) means that
electrons (holes) are donated by the metal substrate to graphene which
becomes $n$-type ($p$-type) doped.

\begin{figure}[!tpb]
\begin{center}
$\begin{array}{cc}
\includegraphics[width=0.6\columnwidth]{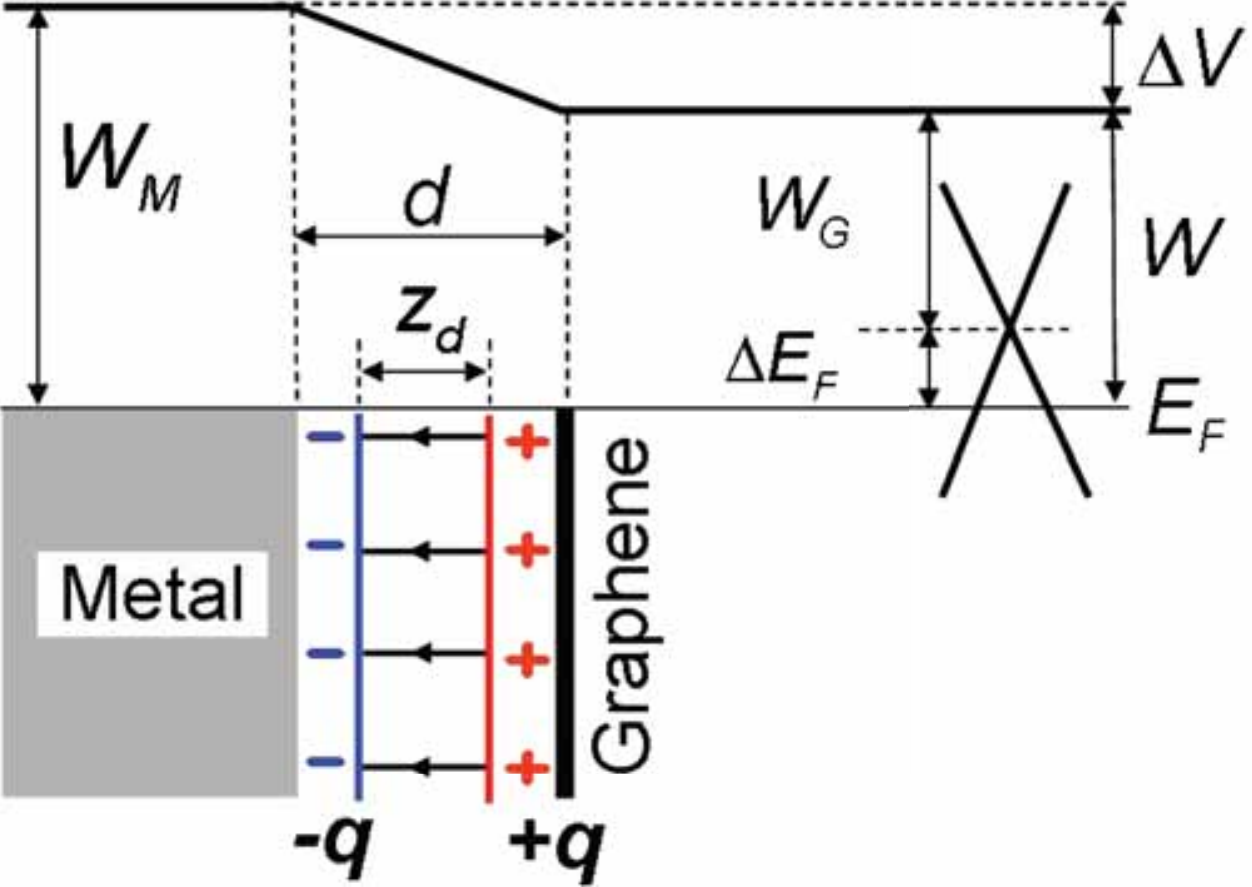} &
\includegraphics[width=0.4\columnwidth]{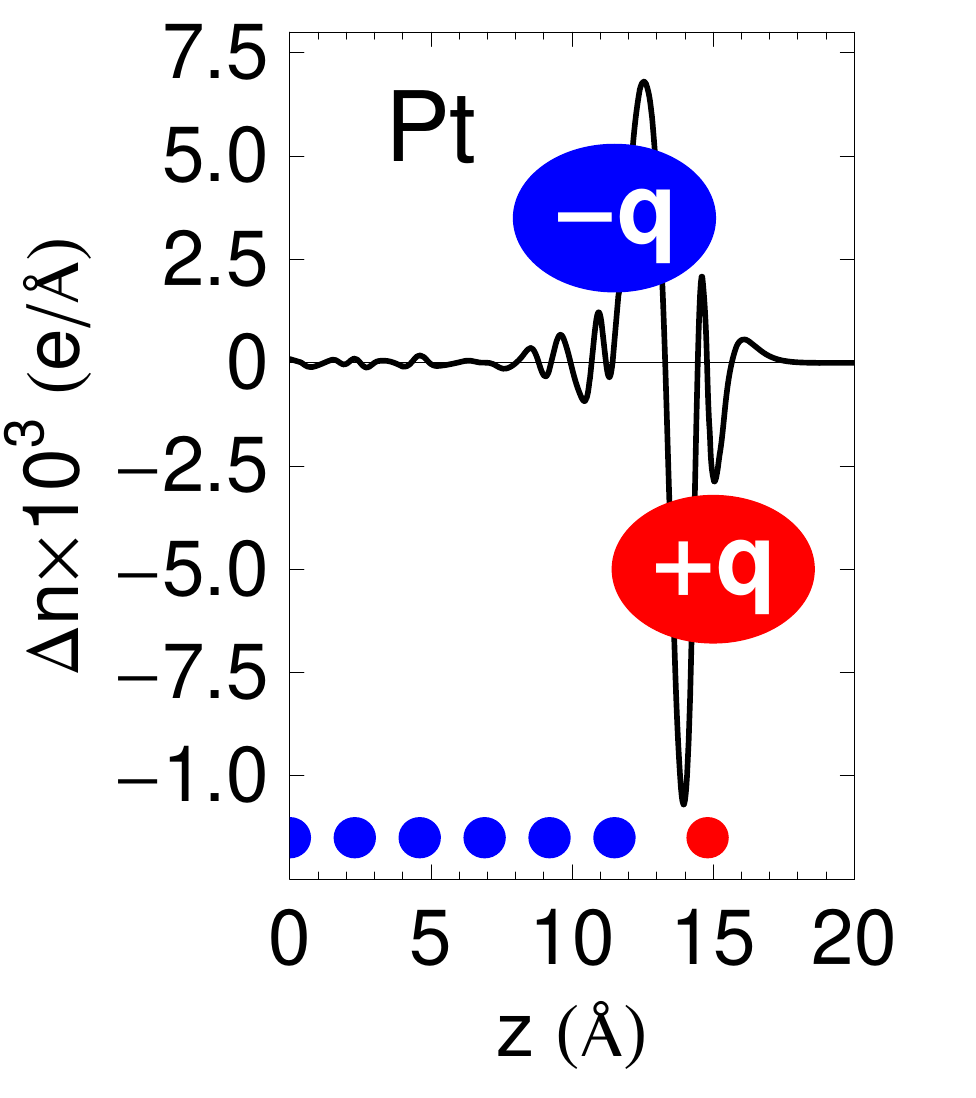}
\end{array}$
\end{center}
\caption{(Color online)
Left: schematic illustration of the parameters used in modeling the
interface dipole and potential step formation at the graphene-metal
interface.
Right: plane-averaged difference electron density
$\Delta n (z) = n_{\rm M \vert G} (z) - n_{\rm M}(z) - n_{\rm G}(z)$
showing the charge displacement upon formation of the graphene-Pt(111)
interface. }
\label{ref:fig4}
\end{figure}

For metal-graphene equilibrium separations, graphene is doped $n$-type
on Al, Ag and Cu, and $p$-type on Au and Pt; the corresponding Fermi
level shifts are plotted in Fig.~\ref{ref:fig3}. Because the work
functions of graphene, $W_{\rm G}$, and of most metal surfaces, $W_{\rm
M}$, differ, as soon as graphene interacts with a metal, electrons are
transferred from one to the other to equilibrate the Fermi levels. A
schematic representation is shown in Fig.~\ref{ref:fig4} for the case
of electron transfer from graphene to the metal. To a good
approximation, the graphene density of states (DOS) is described by
$D(E)=D_{0} |E|$, with $D_{0} = 0.09$/(${\rm eV}^{2}$ unit cell) for
$E$ within 1~eV of the conical points. Since this DOS is much lower
than that of the metal, equilibrium is effectively achieved by moving
the Fermi level in graphene and even a small electron transfer will
shift the Fermi level significantly. A transfer of 0.01 electrons would
lower the Fermi level by 0.47 eV.

This electron transfer results in the formation of an interface dipole
layer and an accompanying potential step $\Delta V$. We can use the
plane-averaged electron densities $n(z)$ to visualize the electron
redistribution $\Delta n (z) = n_{\rm M \vert G} (z) - n_{\rm M}(z) -
n_{\rm G}(z)$ upon formation of the interface. As shown in
Fig.~\ref{ref:fig4}, $\Delta n (z)$ is localized at the interface. The
sign and size of the interface dipole are consistent with the changes
of the metal work function upon adsorption of graphene, see
Table~\ref{ref:tab1}.

Naively one would assume that graphene is doped with electrons if
$W_{\rm G}>W_{\rm M}$ and doped with holes if $W_{\rm G}<W_{\rm M}$.
The crossover point from $n$- to $p$-type doping would then be at
$W_{\rm M}=W_{\rm G}$. The results obtained at the equilibrium
separations of the graphene sheet and the metal surfaces ($d \sim 3.3$
\AA ; see Fig.~\ref{ref:fig3}) show that this is clearly not the case.
Instead, the crossover point lies at $W_{\rm M}-W_{\rm G}=0.9$ eV. Only
when the graphene-metal separation is increased significantly does the
crossover point decrease to its expected value, as illustrated by the
upper curve for $d = 5.0$ \AA\ in Fig.~\ref{ref:fig3}. This clearly
demonstrates that the charge redistribution at the graphene-metal
interface is not only the result of an electron transfer between the
metal and the graphene levels. There is also a contribution from a
metal-graphene chemical interaction.
Such an interaction, which has a significant repulsive contribution, has been
found to play an important role in describing dipole formation when
closed shell atoms and molecules are adsorbed on metal surfaces
\cite{Silva:prl03,Rusu:thesis07}.

\begin{figure}[!tpb]
\begin{center}
\includegraphics[width=1.0\columnwidth]{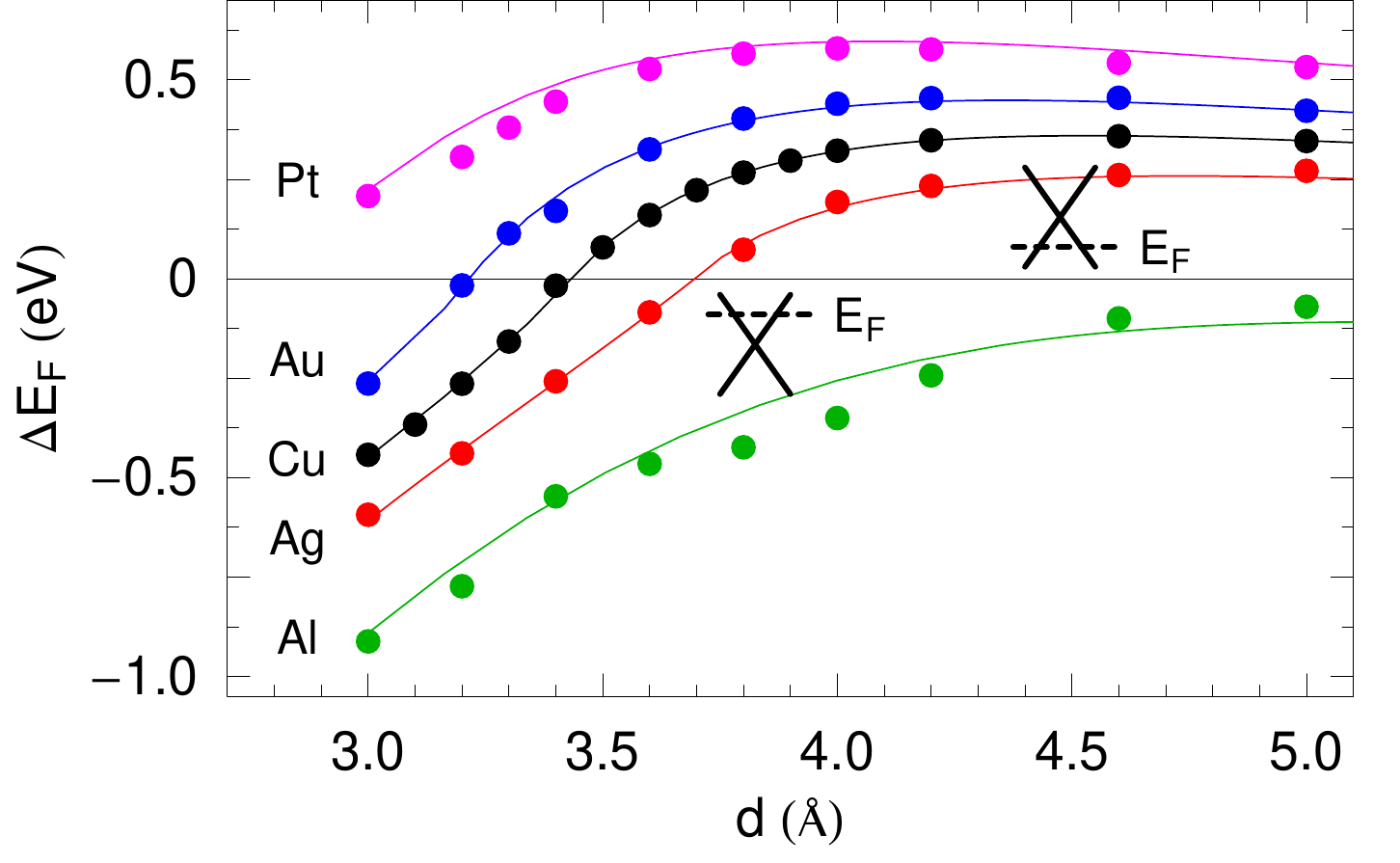}
\end{center}
\caption{(Color online) Fermi level shifts $\Delta E_{\rm F}(d)$ as a function of the graphene-metal surface
distance. The dots give the calculated DFT results, the solid lines give the results obtained from
the model, Eq.~(\ref{eqn1}) \cite{footnote2}.}
\label{ref:fig5}
\end{figure}

The dependence of this interaction on the metal-graphene separation $d$
is mapped out in Fig.~\ref{ref:fig5} in terms of the dependence of the
Fermi level shift $\Delta E_{\rm F}$ on $d$. We use the parameters shown
in Fig.~\ref{ref:fig4} to construct a simple and general model with which
to understand these results. The work function of the graphene-covered
metal is given by $W(d) = W_{\rm M} - \Delta V(d)$ where $\Delta V$
is the potential change generated by the metal-graphene interaction.
The Fermi level shift in graphene is modeled as
$\Delta E_{\rm F}(d) = W(d) - W_{\rm G}$. The key element is
modeling the potential step $\Delta V = \Delta_{\rm tr}(d) + \Delta_{\rm c}(d)$ in
terms of a ``non-interacting'' charge transfer contribution
$\Delta_{\rm tr}$ driven by the difference in work functions and a
contribution $\Delta_{\rm c}$ resulting from the metal-graphene chemical interaction.

The charge transfer contribution is modeled by a plane capacitor model
as indicated in Fig.~\ref{ref:fig4}. $\Delta_{\rm tr}(d)=\alpha N(d)
z_d$ where $\alpha = e^2/\varepsilon_0 A=34.93$ eV/\AA\, with $A =
5.18$ \AA$^2$ the area of the graphene unit cell and $N(d)$ is the
number of electrons (per unit cell) transferred from graphene to the
metal (becoming negative if electrons are transferred from the metal to
graphene). $z_d$ is the effective distance between the charge sheets on
graphene and the metal. $z_d < d$ as most of the charge is located
between the graphene layer and the metal surface as illustrated in
Fig.~\ref{ref:fig4}. We model it as $z_d = d - d_0$ with $d_0$ a
constant.

Integrating the (linear) density of states of graphene yields a simple
relation between $N(d)$ and $\Delta E_{\rm F}(d)$: $N =\pm D_{0} \Delta
E_{\rm F}^2/2$. Using the relations
introduced in the previous two paragraphs we can then express $\Delta
E_{\rm F}(d)$ as
\begin{equation}
\Delta E_{\rm F} (d)\hspace{-0.5mm} =\hspace{-0.5mm} \pm \frac{ \sqrt{1\hspace{-0.5mm} +\hspace{-0.5mm} 2 \alpha D_0 ( d \hspace{-0.5mm} - \hspace{-0.5mm} d_0 ) \vert  W_{\rm M}\hspace{-0.5mm} -\hspace{-0.5mm} W_{\rm G}\hspace{-0.5mm} -\hspace{-0.5mm} \Delta_{\rm c}(d) \vert } \hspace{-0.5mm}- 1 }{ \alpha D_0  ( d\hspace{-0.5mm} -\hspace{-0.5mm} d_0 )},
\label{eqn1}
\end{equation}
where the sign of $\Delta E_{\rm F}$ is given by the sign of $W_{\rm
M}-W_{\rm G}-\Delta_{\rm c}$. The parameters $d_0$ and $\Delta_{\rm
c}(d)$ turn out to depend only very weakly on the metal substrate. If
we fit these quantities to the DFT results for one metal substrate, we
can use them as universal parameters to predict the Fermi level shifts
in graphene for all metal substrates. We use the DFT results obtained
for graphene on Cu (111) to fix $d_0$ and $\Delta_{\rm c}(d)$, see
Ref.~\cite{footnote2}. Only the work function of the clean metal
surface, $W_{\rm M}$, and that of free-standing graphene, $W_{\rm G}$,
are then needed to calculate the Fermi level shift. The accuracy of the
model represented by Eq.~(\ref{eqn1}) is demonstrated in
Figs.~\ref{ref:fig3} and \ref{ref:fig5}. From $\Delta E_{\rm F}$ one
can immediately obtain the work function $W$ of the metal-graphene
system, as well as the sign and concentration of the charge carriers in
graphene, $N$.

The critical metal work function $W_{\rm M}=W_{0}$ where the Fermi
level is at the conical points of graphene, can be obtained from
Eq.~(\ref{eqn1}) for $\Delta E_{\rm F} (d)=0$. It gives $W_{0}(d) =
W_{\rm G} + \Delta_{\rm c}(d)$. The contribution of the chemical
interaction term $\Delta_{\rm c}$ depends strongly on the distance $d$
between graphene and the metal surface. At a large distance $d\gtrsim
4.2$ \AA, $\Delta_{\rm c}\ll 1$ eV and $W_{0}(d) \approx W_{\rm G} =
4.5$ eV, whereas at the equilibrium separation $d_{\rm eq} = 3.3$ \AA,
$\Delta_{\rm c}\approx 0.9$ eV and $W_{0}(d) \approx 5.4$ eV. This
agrees with the DFT results shown in Fig.~\ref{ref:fig3}. The chemical
interaction thus leads to a sizeable potential step at the equilibrium
separation, which is downwards from metal to graphene as indicated in
Fig.~\ref{ref:fig4}. The sign of this step and its insensitivity to the
metal substrate are consistent with its interpretation in terms of an
exchange repulsion between the electrons on graphene and the metal
substrate \cite{Rusu:thesis07}.

In conclusion, we have used DFT calculations to study the doping
of graphene induced by adsorption on metal surfaces and developed a simple
model that takes into account the electron transfer between the
metal and graphene levels driven by the work function difference, as
well as the chemical interaction between graphene and the metal. The
model extends the applicability of the detailed DFT results to the
more complex systems encountered in practical devices and suggests combinations of
metal (strips) to be used to realize $p$-$n$ junctions \cite{Cheianov:prb06}.

We thank Paul Rusu for helpful discussions. This work was financially
supported by ``NanoNed'' (a programme of the Dutch Ministry of Economic
Affairs) and by the ``Nederlandse Organisatie voor Wetenschappelijk
Onderzoek (NWO)'' via ``Chemische Wetenschappen (CW)'' and ``Stichting
voor Fundamenteel Onderzoek der Materie (FOM)''. Computer facilities
were granted by ``Stichting Nationale Computerfaciliteiten (NCF)''.

\end{document}